\documentclass[aps, prd, showpacs, showkeys, preprint]{revtex4}
\usepackage{epsfig}

\begin{document}

\begin{titlepage}
~\vskip1cm

\title{Analytic structure
 in the coupling constant  plane in perturbative QCD}

\author{Irinel Caprini}
\affiliation{National Institute of Physics and Nuclear Engineering,
POB MG 6, Bucharest, R-76900 Romania}
\author{Jan Fischer}
\affiliation{Institute of Physics, Academy of Sciences of the Czech Republic,
CZ-182 21  Prague 8, Czech Republic}

~\vskip1cm

\begin{abstract} We investigate the  analytic structure
 of the Borel-summed perturbative QCD  amplitudes in the complex plane of the coupling  
constant. Using the method of inverse Mellin
 transform, we show that  the prescription dependent 
Borel-Laplace integral can be cast, under some conditions, 
into the form of a dispersion relation in the $a$-plane.
We also discuss some recent works relating resummation prescriptions, 
renormalons and nonperturbative effects, and show that a method proposed 
recently for obtaining QCD nonperturbative condensates from perturbation theory is 
based on  special assumptions about the analytic structure 
in the coupling plane that are not valid in QCD.
 \end{abstract}

\pacs{12.38.Bx, 12.38.Cy, 12.38.Aw}

\keywords{QCD, renormalons, analytic properties}

\maketitle

 \end{titlepage}
 \newpage

\section{Introduction} \label{intro}

 The QCD amplitudes  have a complicated analytic structure
 in the  complex plane of the coupling constant, $\alpha_s=g^2$. As proved in 
 \cite{tHooft, Khuri}, the infinitely many multiparticle branch points at large energies 
 result, via renormalization group invariance, in 
 an accumulation of essential singularities near the origin 
 $\alpha_s=0$. Since in massless QCD the multiparticle hadronic states are 
 generated only by a nonperturbative confinement mechanism, these 
 singularities can show up only  beyond perturbation theory.
 However, the perturbative amplitudes  themselves are expected to have a 
 complicated structure of singularities, due to the fact that the  
 perturbation series is divergent and  Borel non-summable. The presence of the 
 renormalons on the
real axis of the Borel plane induces  singularities of the amplitudes as 
functions of the coupling constant.

 In a recent paper \cite{Lee}, arguments based on analyticity in the coupling 
 complex plane were used to suggest the possibility of calculating genuine 
 nonperturbative quantities, like QCD condensates, from pure perturbation theory. 
 The analytic structure and its connection with infrared renormalons were  further 
 discussed in \cite{Cvetic, Lee1}.
  Motivated by this recent interest in the problem, we investigate in the present 
  work the analytic structure in the complex coupling plane of the Borel-summed 
  amplitudes in perturbative QCD.  We use the mathematical techniques applied in
  \cite{CaNe, CaFi, CaFi1}, which  allow us to express the Borel integral as a
dispersion relation in the coupling plane.   In the last section 
  we shall make a few comments  on the papers \cite{Lee, Cvetic, Lee1}.

\begin{figure}
\includegraphics{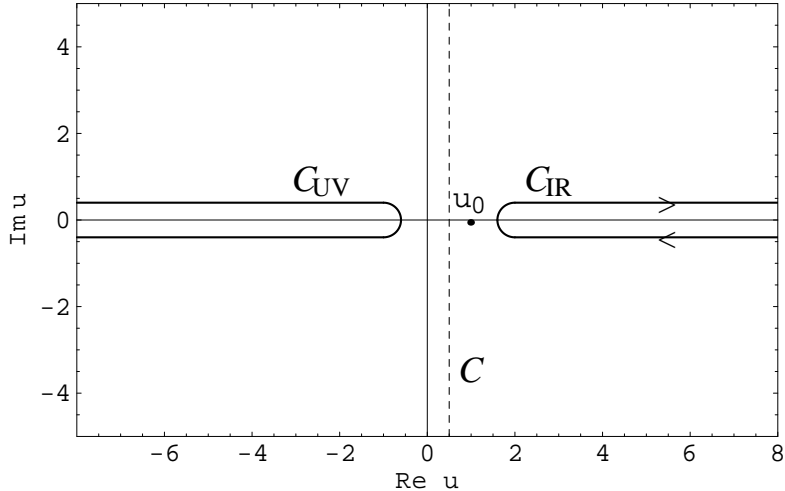}
\caption{The Borel plane for the Adler function}
\label{fig1}
\end{figure}

We consider for illustration the Adler
  function in massless QCD
\begin{equation}\label{Adler} 
{\cal D} =- Q^2 {{\rm d}\Pi(Q^2)\over {\rm d}Q^2}-1\,,
\end{equation}
where $\Pi(Q^2)$ is the current-current correlation function calculated for euclidian arguments
  $Q^2>0$. It is known that the perturbation expansion of ${\cal D}_{PT}(a)$ in powers of the renormalized
 coupling  $a=\alpha_s(Q^2)/\pi$  is divergent and not  Borel summable (see \cite{Beneke} and references therein).
The attempts of performing the summation by a formal Borel-Laplace integral \cite{Hardy} encounter the difficulty
 that this integral is not well-defined. We consider the Borel transform $B(u)$ defined
  in the standard way in terms of the  perturbative coefficients $d_n$ of ${\cal D}$:
\begin{equation}\label{B} 
B(u)=\sum\limits_{n=0}^\infty {d_n\over n!}\, \left({u\over \beta_0}\right)^n\,,
\end{equation}
  where $\beta_0=(33-2 n_f)/12$ is the first  QCD beta-function coefficient with $n_f$
the number of  flavors. It is known,  from the $n!$ large order growth of $d_n$,  that $B(u)$ has
 singularities (ultraviolet  and infrared renormalons) on the real axis of the $u$-plane \cite{Beneke}.
 For the Adler function, the ultraviolet renormalons are placed along the range 
$u\leq -1$ and the  infrared renormalons  along
 $u \geq 2$  (see Fig. \ref{fig1}). Due to the infrared renormalons, 
  the usual Borel-Laplace integral  is not well-defined  and requires
 an  integration prescription. Defining
\begin{equation}\label{pm}   {\cal D}_{PT}^{(\pm)}(a)={1\over \beta_0
}\,\int\limits_{{\cal C}_\pm}\!{\rm e}^{-{u\over \beta_0 a}} \, 
B(u)\,{\rm d}u\,= {1\over \beta_0
}\,\lim\limits_{\epsilon\to 0}\,\int\limits_{0\pm i\epsilon}^{\infty\pm 
i\epsilon}\!{\rm e}^{- {u\over \beta_0 a} } \, B(u)\,{\rm d}u\,,\end{equation}
 one can adopt as prescription, for each value of $a$ with ${\rm Re}\, a>0$,
 either   ${\cal D}_{PT}^{(+)}(a)$ or ${\cal D}_{PT}^{(-)}(a)$, or 
 a linear combination of them, with coefficients $\xi$ and $1-\xi$
such as to correctly reproduce the known 
low-order expansion of ${\cal D}_{PT}(a)$ (which is obtained
by truncating the Taylor expansion (\ref{B}) at a finite order $N$).
 We consider in particular
 the principal value ($PV$) prescription
\begin{equation}\label{pv}
{\cal D}_{PT}^{(PV)}(a)={1\over 2} [{\cal D}_{PT}^{(+)}(a)+ {\cal D}_{PT}^{(-)}(a)]\,.
\end{equation}
Once a prescription is adopted, one has a well-defined  function 
of $a$, different prescriptions yielding different functions. 
In the next section  we shall study the analytic properties of
these functions 
 in the complex $a$-plane. 

\section{Dispersion relations for the Borel summed amplitude}

The  analytic properties of the integrals (\ref{pm}) with 
respect to the variable $a$ can be studied with standard mathematical techniques. 
In the present work, we  use the method of inverse Mellin
transform, applied for the first time in the context of  Borel summation  in QCD in  \cite{Neub, BBB}. 
 For details of the mathematical procedure used  below, we refer to
  \cite{CaNe}, where the same method was applied  for investigating
the  momentum-plane analyticity structure of the Adler function in  the large-$\beta_0$ limit.
 
 The inverse Mellin transform  of the function 
$ B(u)$ is defined by \cite{Tit} 
\begin{equation}\label{wD}
   \widehat w (\tau) = \frac{1}{2\pi i}
   \int\limits_{u_0-i\infty}^{u_0+i\infty}\!
    B(u)\,\tau^{u-1}{\rm d}u\,
\end{equation}
and admits the inverse   relation  
\begin{equation}\label{wDinv}
    B(u) = \int\limits_0^\infty\!
    \widehat w(\tau)\,\tau^{-u}{\rm d}\tau\,,
\end{equation}
which gives $B (u)$ in the strip $-1<\mbox{Re}\,u< 2$  parallel to the imaginary 
axis \cite{Tit}. 
The above relations are valid if the  Borel transform satisfies the 
condition \cite{Tit}
\begin{equation}\label{L2B}
   \frac{1}{2\pi i} \int\limits_{u_0-i\infty}^{u_0+i\infty}\!
   |B(u)|^2 {\rm d}u < \infty \,,
\end{equation}
where $u_0$ is a point located on the real axis, between the branch 
points, $-1< u_0 < 2$ (see Fig.1). This  condition  strongly
 restricts the asymptotic behaviour 
of the Borel transform, and it is not known whether it is obeyed or not in  QCD. 
The condition (\ref{L2B}) is however  satisfied in some particular cases of physical interest.
 One example is  the summation of a chain of diagrams in the large-$\beta_0$ (or large-$n_f$ limit),
which leads to  the Borel transform  \cite{Bene, Broa}:
\begin{equation}\label{Blb0}
    B (u) = \frac{128}{3(2-u)}\,\sum_{k=2}^\infty\,
   \frac{(-1)^k\,k}{\big[ k^2-(1-u)^2 \big]^2} \,,
\end{equation}
 working in the $V$-scheme, where all the exponential factors  are included in the 
definition of the coupling.  As a second example we take the case of a  finite number of renormalons with
 branch-point singularities, in particular
the contribution to $B(u)$ of the leading infrared and ultraviolet renormalons 
\begin{equation}\label{Bfirst}
B(u)= {K\over (1- u/p)^{\nu+1}}+
{K' \over (1+ u/p')^{\nu'+1}}\,,
\end{equation}
 where $p>0$, $p'>0$ and the constants $K$ and $K'$ represent
 the strength of the
corresponding singularities. We note that for the Adler function  $p=2$, $p'=1$, 
and the exponents $\nu$ and $\nu'$ were calculated in  \cite{Muel}  
and  \cite{BBK}, respectively.

The function $\widehat w(\tau)$ defined in (\ref{wD}) can be calculated
 by closing the integration contour along a semi-circle at
infinity in the $u$ plane and applying the theorem of residues. For $|\tau|<1$ the contribution 
of the semi-circle at infinity vanishes if the contour is closed in the 
right half of the $u$ plane, while for $|\tau|>1$ the contour must be
closed in the left half plane. Therefore one obtains different 
expressions for the distribution function at
$|\tau|<1$ and $|\tau|>1$. By Cauchy's theorem these functions,  which we denote by 
$\widehat w_<(\tau)$
 and $\widehat w_>(\tau)$, have the representations	
\begin{eqnarray}\label{wD1}
   \widehat w_<(\tau) &=& \frac{1}{2\pi i}\int\limits_{{\cal C}_{\rm IR}}\!
   B(u)\,\tau^{u-1} {\rm d}u \,,\\
 \widehat w_>(\tau) &=& \frac{1}{2\pi i}\int\limits_{{\cal C}_{\rm UV}}\!
   B(u)\,\tau^{u-1} {\rm d}u \,,\nonumber
\end{eqnarray}
  the  integration (Hankel) contours ${\cal C}_{\rm IR}$ and
   ${\cal C}_{\rm UV}$ being indicated in Fig. 1.

The inverse Mellin transform   $\widehat w$ was 
 calculated in Ref. \cite{Neub} for the Adler function in the
 large-$\beta_0$ limit,  when the Borel transform has the expression 
(\ref{Blb0}). In this case 
\begin{eqnarray}\label{wDfun}
 \widehat w_<(\tau)&=& \frac{32}{3} \left\{ \tau\left(
    \frac74 - \ln\tau \right) + (1+\tau)\Big[ L_2(-\tau) + \ln\tau
    \ln(1+\tau) \Big] \right\} \,,\\
  \widehat w_>(\tau)&=& \frac{32}{3} \left\{ 1 + \ln\tau
    + \left( \frac34 + \frac12 \ln\tau \right) \frac{1}{\tau}
    + (1+\tau)\Big[ L_2(-\tau^{-1}) - \ln\tau \ln(1+\tau^{-1}) \Big]
    \right\},  \nonumber
\end{eqnarray}
where $L_2(x)=-\int_0^x {{\rm d}t\over t}\ln(1-t)$ is the dilogarithm. The
 physical interpretation of  $\widehat w$ as the distribution 
of the internal gluon virtualities in Feynman diagrams was also  pointed 
out in \cite{Neub}.

The function $\widehat w$ can be also calculated explicitly  for a finite number of 
renormalons with
 branch-point singularities. For instance, for the Borel transform written in (\ref{Bfirst}) we obtain  \cite{Bate}:
\begin{eqnarray}\label{wfirst}
\widehat w_<(\tau) &=&  {K\over \Gamma(\nu+1)} \,p^{\nu+1} \,\tau^{p-1} \,
(-\ln \tau)^{\nu}\,,              \\
\widehat w_>(\tau) &=&  {K'\over \Gamma(\nu'+1)} \, (p')^{\nu'+1}\,
\frac{1}{\tau^{p'+1}} \,(\ln \tau)^{\nu'}\,. \nonumber
\end{eqnarray}

We now proceed to the evaluation of  the  
integrals  (\ref{pm}), taking  $a$  in the right half plane,  ${\rm Re}\, a>0$, where
we assume that they converge.
Our aim is to obtain a representation of ${\cal D}_{PT}^{(\pm)}(a)$  in terms of the 
inverse Mellin transform $\widehat w$. To this end we rotate the integration contours ${\cal C}_\pm$,
whitout crossing singularities, up  to a line parallel to the imaginary axis, 
where the representation (\ref{wDinv}) is valid. It is easy to check that  for  $a$  in the upper half plane
  (${\rm Im}\, a>0$), the contribution of 
the quarter of the circle at infinity vanishes if the rotation is performed in the upper half of the
$u$-plane.
  Using the representation (\ref{wDinv}), valid along the imaginary axis,
 and performing  the integral with respect to 
$u$ (for details see \cite{CaNe}), we obtain
 \begin{equation}\label{Dp1}
{\cal D}_{PT}^{(+)}(a)=\frac{1}{\beta_0} \int\limits_0^{\infty} {\widehat w(\tau)\over 
{1\over \beta_0 a}+ \ln\tau} {\rm d} \tau \,,\quad {\rm Im}\, a >0\,.
\end{equation}
Similarly,  for  $a$  in the lower half plane
  (${\rm Im}\, a<0$)  the  integration axis in the expression of ${\cal D}_{PT}^{(-)}(a)$ 
can be rotated  in the lower half of the $u$-plane, up to the negative imaginary axis, leading to
 \begin{equation}\label{Dm1}
{\cal D}_{PT}^{(-)}(a)=\frac{1}{\beta_0} \int\limits_0^{\infty} {\widehat w(\tau)\over {1\over \beta_0 a}+ \ln\tau} 
{\rm d} \tau\,,\quad {\rm Im}\, a <0\,,
\end{equation}
 We notice further,  recalling the definition (\ref{pm}), that the first relation (\ref{wD1})  can be  expressed as 
\begin{equation}\label{dif}
{\cal D}_{PT}^{(+)}(a)={\cal D}_{PT}^{(-)}(a)+ 2 i\,\sigma_<(a)\,
\end{equation} where we introduced the notation
 \begin{equation}\label{sigma}
\sigma(a)= {\pi\over \beta_0} [\tau  \widehat w (\tau)]_{\tau=e^{-{1\over \beta_0 a}}}\,.
\end{equation}
In particular, using the expression (\ref{wfirst}) of 
  $\widehat w$,  we obtain for the leading renormalons:
\begin{eqnarray}\label{sigma1}
\sigma_<(a)&=& K\, { 2^{\nu+1} \over\Gamma(\nu+1)} \, {\pi \over  \beta_0 }\,
e^{-{2\over \beta_0 a}}\, (\beta_0 a)^{-\nu}\,,\\
\sigma_>(a)&=& K'\, { 1  \over\Gamma(\nu'+1)} \, {\pi \over  \beta_0 }\,
e^{1\over \beta_0 a}\, (-\beta_0 a)^{-\nu'}\,.\nonumber\end{eqnarray} 
 It is important to emphasize that the relation (\ref{dif}) is valid for  ${\rm Re}\, a>0$ (or equivalently for
 $|\tau|<1$),  i.e. in the whole right-half of the $a$-plane.

The relations  (\ref{Dp1}),  (\ref{Dm1}) and (\ref{dif}) are the basis the 
derivation of the dispersion relations
given below. We first note that, unlike the original representations (\ref{pm}) which converge
only for  ${\rm Re}\, a >0$,
the representations (\ref{Dp1}) and  (\ref{Dm1})  can be analytically continued
 in the corresponding upper (lower) half of the complex $a$ plane, outside the real axis. 
 Moreover, it is easy to convert them  into a 
 dispersion representation in the variable $a$. We first split the integral 
  in two integrals, one from 0 to 1 (where $\widehat w=\widehat w_<$), and the 
  other from 1 to $\infty$ (where $\widehat w=\widehat w_>$), and perform 
  in each interval the change of variable
 \begin{equation}\label{change}
\ln \tau=-{1\over \beta_0 a'}\,,\quad {{\rm d}\tau\over\tau}={{\rm d} a'\over \beta_0 (a')^2}\,.
\end{equation}
  Using the relations  (\ref{Dp1})-(\ref{Dm1}) thus transformed, together with 
(\ref{dif})  and (\ref{sigma}),   we  finally express  ${\cal D}_{PT}^{(+)}(a)$ as
 \begin{eqnarray}\label{DRap}
{\cal D}_{PT}^{(+)}(a)&=&{a  \over \pi} \int\limits_0^\infty
{ \sigma_<(a')\, {\rm d} a'  \over a'(a'-a)}+{ a  \over \pi} \int\limits^0_{-\infty} 
{ \sigma_>(a')\,{\rm d} a' \over a'(a'-a)},\quad {\rm Im}\, a>0\,,\\
{\cal D}_{PT}^{(+)}(a)&=&{ a \over \pi}  \int\limits_0^\infty
{ \sigma_<(a')\, {\rm d} a'  \over a'(a'-a)}+{ a \over \pi}  \int\limits^0_{-\infty}
 { \sigma_>(a')\,{\rm d} a' \over a'(a'-a)}+
2i   \sigma_<(a),\quad {\rm Im}\, a<0\,.\nonumber
\end{eqnarray}
From the definition (\ref{sigma}) and the properties of 
 the inverse Mellin transform $\widehat w$, it follows that the function 
 $\sigma_<(a)$ can be analytically continued in the complex $a$-plane. This property is seen 
explicitly in the case of one infrared renormalon in (\ref{sigma1}).
Thus, the expressions (\ref{DRap})  are analytic functions in 
the upper (lower) half of the complex $a$-plane, outside the real axis. From (\ref{sigma1}) it is seen that 
 $\sigma_>(a)$ is real for $a<0$, while  $\sigma_<(a)$ is real for $a>0$. Therefore, the   spectral functions
of the dispersion integrals are real.

Taken together, the dispersion relations
 (\ref{DRap}) define a  single  analytic function, ${\cal D}_{PT}^{(+)}(a)$, in the whole cut $a$-plane. 
  This function  may have a  discontinuity across the positive axis, due the Cauchy 
  dispersion integral along $a > 0$ and  the additional  term $2 i \sigma_<(a)$ in the second relation.
 A simple calculation shows however that
  \begin{equation}\label{DRareal}
\lim_{\epsilon \rightarrow 0_+}\, {\cal D}_{PT}^{(+)}(a\pm i\epsilon)={ a  \over \pi} {\rm P} \int\limits_0^\infty
{  \sigma_<(a')\, {\rm d} a'  \over a'(a'-a)}+ { a  
\over \pi}\int\limits^0_{-\infty}
 { \sigma_>(a')\,  {\rm d} a' \over a'(a'-a)}+ i \sigma_<(a),\quad a>0\,,
\end{equation}
where P denotes the Cauchy principal value.
This relation shows that the function ${\cal D}_{PT}^{(+)}(a)$ is well defined along the 
positive axis, but has there
an unphysical imaginary part equal to
  \begin{equation}\label{Im}
{\rm Im}\, {\cal D}_{PT}^{(+)}(a)= \sigma_<(a)\,,\quad a>0\,.
\end{equation}
From (\ref{DRap}) it follows  that  ${\cal D}_{PT}^{(+)}(a)$ has actually a discontinuity along the negative axis, given by
 \begin{equation}\label{Discnega}
 {\cal D}_{PT}^{(+)}(a+ i \epsilon) -
 {\cal D}_{PT}^{(+)}(a- i \epsilon)= 2i\,[\sigma_>(a) - \sigma_<(a-i\epsilon)] \,,\quad a<0\,,
\end{equation}
in terms of the real spectral function $\sigma_>$ and the analytic continuation of $\sigma_<$
up to the lower edge of the negative semiaxis (where, as seen from (\ref{sigma1}), it is complex).

 The function  ${\cal D}_{PT}^{(-)}(a)$ satisfies a  dispersion relation similar to (\ref{DRap}),
 with an additional term $- 2i   \sigma_<(a)$ in the right hand side, as follows from (\ref{dif}).
As above, one can show that  ${\cal D}_{PT}^{(-)}(a)$ is well-defined for $a>0$, but it assumes there complex values.
 The unphysical  imaginary part  is eliminated if we adopt 
the principal value prescription defined in (\ref{pv}), for which we obtain: 
 \begin{eqnarray}\label{DRppv}
{\cal D}_{PT}^{(PV)}(a)&=&{ a  \over \pi}  \int\limits_0^\infty
{\sigma_<(a')\, {\rm d} a'  \over a'(a'-a)}+{ a  \over \pi}  \int\limits^0_{-\infty}
 {\sigma_>(a')\,{\rm d} a' \over a'(a'-a)}- i \sigma_<(a) \,,\quad {\rm Im}\, a>0\\
{\cal D}_{PT}^{(PV)}(a)&=&{ a  \over \pi}  \int\limits_0^\infty
{\sigma_<(a')\, {\rm d} a'  \over a'(a'-a)}+{ a  \over \pi}  \int\limits^0_{-\infty}
 {\sigma_>(a')\,{\rm d} a' \over a'(a'-a)}+i \sigma_<(a) \,,\quad{\rm Im}\, a<0.\nonumber
\end{eqnarray}
For $a$ on the positive semiaxis, is easy to check that
 \begin{equation}\label{DRpv}
\lim_{\epsilon \rightarrow 0_+}\,{\cal D}_{PT}^{(PV)}(a+i\epsilon)={ a  \over \pi} P \int\limits_0^\infty
{\sigma_<(a') {\rm d} a'  \over a'(a'-a)}+{ a  \over \pi}  \int\limits^0_{-\infty}
 {\sigma_>(a') \,{\rm d} a' \over a'(a'-a)}\,, \quad a>0\,.
\end{equation}
 Therefore, ${\cal D}_{PT}^{(PV)}(a)$  is well-defined and real
 on the positive real semiaxis, as required by general principles. 
 The expressions 
 (\ref{DRppv}) 
  define an analytic function  which satisfies the reality condition $
{\cal D}_{PT}^{(PV)}(a^*)=[{\cal D}_{PT}^{(PV)}(a)]^*$ in the whole complex $a$-plane cut along
 the negative semiaxis, where it has a  discontinuity
\begin{equation}\label{disc}
{\cal D}_{PT}^{(PV)}(a+ i\epsilon)-{\cal D}_{PT}^{(PV)}(a-i\epsilon)= 2 i [\sigma_>(a)- {\rm Re }\,\sigma_>(a)]\,,\quad a<0.\end{equation}
 As argued in \cite{CaNe}, this discontinuity vanishes only under strong restrictions on
 the asymptotic behavior of $B(u)$, which are satisfied neither in the simple cases considered here, nor,
 most probably, in full QCD: namely, the $L^2$ condition (\ref{L2B}) must be satisfied 
not only by $B(u)$, but also by the product $B(u)\,\sin\pi u$ (for technical details, see Ref. \cite{CaNe}).

We point out that the imaginary part 
 of  ${\cal D}_{PT}^{(PV)}(a)$  along the positive semiaxis vanished due to a 
precise cancellation of the imaginary part of the integrals and  the last terms in  Eqs. (\ref{DRppv}). 
It is easy to check that for a general linear combination of ${\cal D}_{PT}^{(\pm)}$, with coefficients
 $\xi$ and $1-\xi$  as discussed above (\ref{pv}), 
  this cancellation no longer holds. Therefore, the principal value prescription
is the most suitable choice if one wants to preserve in perturbative QCD the  
analytic  properties of the true amplitudes.

In  this section, we obtained the nontrivial result that the inverse Mellin transform can be used to derive
from  (\ref{pm})  dispersion relations for the  perturbative amplitudes  ${\cal D}_{PT}^{(\pm})(a)$, 
which allow an analytic continuation of the Borel integral into the left-hand half-plane ${\rm Re}\,a<0$
and  explicitly exhibit the singularities and the discontinuities across the cuts.
 
 The results obtained here,  besides expressing  
the Borel integral (\ref{pm}) in the more suitable form of a  dispersion relation,  will be useful in discussing
 the validity of some assumptions on analyticity in the $a$ plane made in the literature,
 as we  show in the next section.

\section{Comments}
In this section we shall make a few comments on the recent papers \cite{Lee, Cvetic, Lee1}, related to the present work. 

In Ref. \cite{Cvetic} the author considers, in 
 the case of a single infrared renormalon,  the problem of removing  the unphysical imaginary part
of  ${\cal D}_{PT}^{(+)}(a)$ by  subtracting a suitable regularization function from it.
 Our results in this particular case are consistent with \cite{Cvetic}.
Indeed,  Eqs. (\ref{DRap}) and  (\ref{DRppv})  show that the function  ${\cal D}_{PT}^{(PV)}(a)$  is 
obtained  by subtracting from
 ${\cal D}_{PT}^{(+)}(a)$  the function
  \begin{equation}\label{Delta}
\Delta(a)= i \sigma_<(a)\,,
\end{equation} which, using (\ref{sigma1}), can be written for $a=|a| e^{i\psi}$ as 
\begin{equation}\label{Delta2}
\Delta(a)=K\, {2^{\nu+1}  \over \Gamma(\nu+1)}\, \left({ \pi\over\beta_0}\right)\,
e^{-{2\over \beta_0 a}}\, |\beta_0 a|^{-\nu} [\sin \psi\nu +i \cos\psi\nu] \,.
\end{equation} 
This expression coincides with  Eq. (17) of 
\cite{Cvetic}, derived using arguments based on regularity with respect to the parameter $\nu$.

 The analytic structure in the coupling constant plane was recently considered 
 also in Ref. \cite{Lee}, where arguments based on analyticity were used
in support of the  claim that the QCD  condensates can be 
determined  using the coefficients of the perturbation series.
In what follows we shall briefly analyse the validity of this claim.

The author of \cite{Lee}  uses the analogy with some
semiclassical models \cite{Zinn}, where a specific contribution (for instance, 
multi-instantons)  is regular for negative couplings  and can be obtained
 for positive couplings by analytic continuation. 
 He invokes the heuristic argument  according to which a perturbation 
series with a sign-nonalternating, $n!$ large order  behavior, can be summed at $a<0$  by a Borel 
integral along the negative axis in the $u$-plane,  where it becomes
 sign alternating.  We note however that sign alternation does not
 necessarily imply Borel summability: 
this property is violated if a nonalternating component is present,
however negligible it may be in comparison with a strong sign-alternating 
component of the series. This is exactly the situation in QCD: due to  the ultraviolet 
renormalons, the Borel integral along the negative $u$-axis is not well defined.

Another conjecture adopted in \cite{Lee} is related to the  nonperturbative amplitude ${\cal D}_{NP}(a)$, 
which must be added to  ${\cal D}_{PT}(a)$   in order to
compensate its unphysical imaginary part:
\begin{equation}\label{cond}
{\rm Im}\, {\cal D}_{NP}(a+ i\epsilon)+{\rm Im}\, {\cal D}_{PT}(a+ i\epsilon) = 0,\quad a>0\,.
 \end{equation}
 According to current interpretations \cite{Beneke}, the cancellation is expected to occur if both terms 
in  the above relation are calculated with the same prescription.
In \cite{Lee} the author supplements (\ref{cond}) by a specific assumption
about  the nonperturbative  amplitude,  taking it of the form 
\begin{equation}\label{DNP0}
{\cal D}_{NP}(a)= C\,e^{-{2\over \beta_0 a}} (-\beta_0\,a)^{-\nu}\,,
 \end{equation}
where $\nu$ is the branch-point exponent of  the infrared renormalon in 
 (\ref{Bfirst})  and  $C$ is related to the gluon condensate (we use the parametrization
 given in \cite{Lee1}, with our notation $a=\alpha_s/\pi$).
This expression is  regular for negative $a$, having
 a discontinuity {\em only} on the positive semiaxis in the $a$-plane, where
\begin{equation}\label{DNP}
{\cal D}_{NP}(a\pm  i\epsilon)= C\, e^{-{2\over \beta_0 a}}\, 
(\beta_0\,a)^{-\nu}\,[\cos\pi\nu \mp i \sin\pi\nu] \,, \quad a>0\,.
 \end{equation}
 Using this relation and the imaginary part of the perturbative amplitude
given in (\ref{Im}) and
  (\ref{sigma1}),  condition  (\ref{cond}) gives the relation
\begin{equation}\label{residue}
K=C\,{1 \over 2^{\nu+1}}\,{\beta_0\over  \Gamma(-\nu)}\,,
 \end{equation}
 from which, according to \cite{Lee}, one could obtain  the nonperturbative parameter $C$
 (the gluon condensate), using the strength  $K$ of the infrared renormalon
 computed from the perturbation series. 

  The relation (\ref{residue}) implies that $K$  vanishes
  for nonnegative  integer $\nu$, when the renormalon 
 $1/(1-u/2)^{1+\nu}$ becomes a pole.
It is known however  that poles are actually  obtained from some  chains of 
Feynman diagrams, in the large-$\beta_0$ (or large-$n_f$) limit \cite{Bene, Broa}.
 In  \cite{Lee1}  the author discusses 
 this limit, taking  the exponent $\nu$   of the form $\nu=\kappa+\chi/ \beta_0+\ldots$,
with $\kappa$ an integer. In this limit, the factor $\Gamma(-\nu)\sim \beta_0/\chi$ in the denominator of (\ref{residue})
 is compensated by the factor $\beta_0$ in the numerator, leading to  a finite nonzero limit for $K$, if
  $C$ tends to a nonzero constant in the large-$\beta_0$ limit. But then  Eq. (\ref{DNP})  implies that the real part of  the nonperturbative amplitude
 $ {\cal D}_{NP}$  is  nonvanishing in this limit  (we keep, following \cite{Lee1},
 the product $\beta_0 a$ constant, which is legitimate,
as seen  in particular in the one-loop expression $a={1 \over \beta_0 \ln Q^2/ \Lambda^2}$).
On the other hand, 
 the  perturbative amplitude  $ {\cal D}_{PT}$ is subleading for large $\beta_0$, due to
 the  factor $1/\beta_{0}$ in front of the Laplace-Borel  integral  
 (\ref{pm}). A choice of the constant $C$ of the same order, i.e. $C\sim c/\beta_0$,
 would imply from (\ref{residue}) that
$K$ vanishes when $\beta_0\to\infty$.
So, it follows  from  \cite{Lee, Lee1} that in the large-$\beta_{0}$ limit
 either the total Adler function is  dominated  by the real part of the nonperturbative amplitude, 
or the renormalon residue
 vanishes.
 
  The  above  implications  depend however, in a  crucial way, on the conjecture
made in \cite{Lee}  that
 the only singularities of the nonperturbative amplitude are along the positive 
semiaxis. Actually, singularities
along the negative semiaxis (produced by
 the ultraviolet renormalons in the OPE coefficients) cannot be excluded.   In \cite{Lee} it is claimed
 that the influence of the ultraviolet 
 renormalons  can be suppressed by an appropriate conformal 
 mapping \cite{Muel}. A numerical suppression, however, does not imply removal 
of the corresponding singularities. As is shown in \cite{CaFi1}, even if $B(u)$ is expanded in
powers of the  optimal conformal mapping of the Borel plane \cite{CaFi2}, 
the function ${\cal D}_{PT}(a)$ does have
 singularities  along the negative semiaxis. Therefore, the general expression of 
${\cal D}_{NP}(a)$  may  have singularities
 both for $a>0$ and for  $a<0$. To give an example, we add to (\ref{DNP}) 
the analogous branch-point term  suggested by   Eq.(\ref{Delta}), taking
 \begin{equation}\label{DNP1}
 {\cal D}_{NP}(a)=  C \, e^{-{2\over \beta_0 a}} (-\beta_0\, a)^{-\nu} + C'\, e^{-{2\over \beta_0 a}} (\beta_0\,a)^{-\nu}\,,
 \end{equation}
where $C$ and $C'$ depend in general on $\nu$ and   $C'=C_R'+iC_I'$ is  complex 
  (as the condition (\ref{cond}) holds in a definite prescription, we specifically refer here to
the prescription leading to ${\cal D}_{PT}^{(+)}$). Then   condition (\ref{cond}) gives
 \begin{equation}\label{cond1}
 - C\, \sin\pi\nu + C_I' + K\, { 2^{\nu+1} \over\Gamma(\nu+1)} \, {\pi \over  \beta_0 }\,   =0\,,
 \end{equation}
which does not  necessarily imply that  $K$ vanishes when  $\nu$ is a nonnegative integer.
 Moreover,  the real part of the nonperturbative amplitude (\ref{DNP1}) 
contains the additional parameter $C_R'$  not specified by the
 condition (\ref{cond1}), which, combined with the first term in the
 expression (\ref{DNP1})  can 
make  ${\rm Re}\, {\cal D}_{NP}$  subleading in the large-$\beta_0$ limit.
 This counterexample disproves the conclusions (that appear to be 
inherent in the method of \cite{Lee, Lee1})  that the renormalon residue vanishes when the singularity is a pole, 
or the amplitude is
dominated  in the large-$\beta_0$ limit by the nonperturbative part. Even more important, however, 
is the fact that the real 
part of the nonperturbative amplitude cannot be determined from the requirement (\ref{cond})
 and analyticity arguments. The possibility, advocated in \cite{Lee}, of obtaining  the gluon 
condensate from the perturbation series fails to work  here.

In the present paper, by using the inverse Mellin transform,  we showed that the
   Borel-summed QCD perturbative amplitudes (\ref{pm}) satisfy 
  dispersion relations 
 which explicitly exhibit  the singularities in the complex $a$-plane and the discontinuities across the cuts, and
 also allow the analytic continuation of 
the Borel integral into the left-hand half-plane ${\rm Re}\,a<0$.
 We saw that the  derivation  relies on some special hypotheses
about the properties of the Borel transform in perturbative QCD:  we namely
 assumed that the Borel transform  has no
 singularities in the complex $u$-plane except for branch-points on the real axis, with a holomorphy
gap around the origin, its asymptotic behavior being such that the 
inequality (\ref{L2B}) holds and the inverse Mellin transform (\ref{wD}) exists. 
We point out that the holomorphy of $B(u)$ in the double-cut $u$-plane  is expected on general grounds to hold in
renormalisable field theories
 \cite{Pari, Muel1}.  The validity of this condition in QCD is presently
almost universally adopted as a plausible assumption.
 As for the asymptotic condition,  its  validity in full QCD is not guaranteed, but 
it is satisfied in the large-$\beta_0$ limit and for a finite number of 
renormalons with branch-point singularities.

In the second part of the paper we made some comments
 on the papers \cite{Lee, Cvetic, Lee1}, where arguments based on analyticity
were used to discuss the connection between renormalons and nonperturbative quantities in QCD. 
 We  investigated some specific conjectures made in \cite{Lee, Lee1} and showed that they are in conflict with the 
 analyticity properties related to the ultraviolet and infrared renormalons, which
 throws  doubts on the method of calculating  QCD condensates from perturbative expansions.
\begin{acknowledgments}
We thank Prof. Jiri Chyla  for many interesting and stimulating discussions on topics related to this work. 
 We also thank Dr. T. Lee for useful correspondence.
One of us (I.C.) thanks  Prof. S. Randjbar-Daemi and  Prof. C. Verzegnassi for hospitality at  ``Abdus Salam'' ICTP and INFN Sezione di Trieste.
 This work was  supported
 by the CERES Program of Romanian MECT, by the Romanian Academy under
 Grant Nr. 23/2003, and by
 the Ministry of Industry and Trade of the Czech Republic, Project RP-4210/69.
\end{acknowledgments}

\end{document}